\documentclass{aa}  

\usepackage{graphicx}
\usepackage{txfonts}
\usepackage{xcolor}
\usepackage{url}

\begin{document}

\title{A Search for Water Ice in the Coma of Interstellar Object 2I/Borisov}

\author{Bin Yang\inst{1}\and 
Michael S. P. Kelley\inst{2}\and 
Karen J. Meech\inst{3}\and 
Jacqueline V. Keane\inst{3}\and 
Silvia Protopapa\inst{4}\and
Schelte J. Bus\inst{3}}

\institute{European Southern Observatory, Alonso de Cordova 3107, Vitacura, Santiago, Chile  \and
Department of Astronomy, University of Maryland, College Park, MD 20742-2421, USA  \and
Institute for Astronomy, 2680 Woodlawn Drive, Honolulu, HI 96822 USA  \and
Southwest Research Institute, Boulder, CO 80302, USA}

 
  \abstract
   {}
   {Interstellar Objects (ISO) passing through our Solar System offer a rare opportunity to probe the physical and chemical processes involved in solid body and planet formation in extrasolar systems. The main objective of our study is to search for diagnostic absorption features of water ice in the near infrared (NIR) spectrum of the second interstellar object 2I/2019 Q4 (Borisov) and compare its ice features to those of the Solar system icy objects.}
   {We observed 2I in the NIR on three separate occasions. The first observation was made on 2019 September 19 UT using the SpeX spectrograph at the 3-m IRTF and again on September 24 UT with the GNIRS spectrograph at the 8-m GEMINI telescope and the last observation was made on October 09 UT with IRTF.}
   {The spectra obtained from all three nights appear featureless. No absorption features associated with water ice are detected. Spectral modeling suggests that water grains, if present, comprise no more than 10\% of the coma cross-section. The comet consistently exhibits a red D-type like spectrum with a spectral slope of about 6\% per 1000~\AA, which is similar to that of 1I/'Oumuamua and is comparable to Solar system comets.}
   {}

 \keywords{comets: individual (2I) --- 
comets: general}

 \maketitle
%

\section{Introduction} 
\label{sec:intro}
Understanding how planetary systems formed from protoplanetary disks is one of the hottest topics in astronomy today. Solar system formation models show that as giant planets form, large numbers of planetesimals are ejected into the interstellar medium \citep{Walsh2011, Shannon2015, Raymond2018}. Consequently, a large number of extrasolar planetesimals must be traversing through interstellar space, and some eventually pass through our Solar system. These interstellar objects (ISOs) are expected to be icy (i.e.\ comet-like) with only a small fraction being rocky objects \citep{meech2016, engelhardt2017}. While the mass distribution and chemical abundances within our Solar system are well studied, the knowledge of other planetary systems is highly limited. Whether the Solar system is typical of planetary systems in general is an enduring question. ISOs provide a rare opportunity to study the details of planet-building processes in extrasolar planetary systems by delivering material to our Solar system for in-depth observations.

The first ISO, 1I/'Oumuamua, was discovered on 2017 October 19 and an extensive 2.5 month follow up campaign was conducted \citep{meech2017}. Astrometric measurements from the ground and HST showed a 30$\sigma$ detection of a radial non-gravitational acceleration varying as $\sim$ r$^{-2}$. Several hypotheses to explain this acceleration were tested and ruled out. The only feasible explanation for the acceleration requires a modest level of cometary outgassing \citep{micheli2018}. Therefore, 1I is likely an icy body in spite of its asteroidal appearance. However, spectroscopic observations from 0.3 to 2.5 $\mu$m did not reveal any diagnostic absorption features of water ice or hydrated minerals \citep{Fitzsimmons2018}. The measured spectral slopes of 1I in the optical wavelength region vary from 10\%/100nm to 17\%/100nm \citep{Ye2017, Fitzsimmons2018}, which are interpreted as low-albedo organic compounds that have undergone exposure to cosmic ray \citep{Fitzsimmons2018}.

C/2019 Q4 (Borisov) was discovered on 2019 August 30 and quickly found to be on a hyperbolic orbit (MPEC R106; Sep. 11, 2019) suggesting an interstellar origin. A dynamical analysis indicates that this eccentricity was not caused by an interaction with a giant planet, since the high eccentricity remains when the orbit is integrated backward to a time before its entrance into the Solar system. Observational effects, or cometary outgassing, are also insufficient to explain the extreme orbital eccentricity of the object, making C/2019 Q4 the second known ISO, designated as 2I/Borisov (MPEC S71; 2019 Sep. 24). Unlike 1I/'Oumuamua, 2I displays a dust coma at 2.8 au from the Sun, typical of Solar system comets at this distance. \cite{Fitzsimmons2019} reported the first detection of the CN (0-0) emission band at 3880$\AA$ in the optical spectrum of 2I. \cite{Opitom2019} confirmed the CN detection but concluded that 2I is highly depleted in C$_2$. Current observations of 2I suggest that it contains volatiles and its composition may be similar to Solar system carbon-chain depleted comets. 

Water ice, one of the planetary-building blocks, is of great importance to the formation of planets \citep{AHearn2011}. The presence of water ice enhances solid surface density and increases sticking efficiency, which in turn, catalyzes the rapid formation of planetesimals and decreases the timescale of giant planet core accretion \citep{Min2016}. Ices are sensitive to changes induced by thermal and radiation processing, so their characteristics and abundances can provide central clues to those aspects of planetary heritage \citep{mumma2011}. The solid state features of water ice are influenced by its formation temperature and thermal history, and references therein]{Grundy1998}. Water ice is commonly detected via the distinctive infrared absorption bands at 1.5 and 2.0 $\mu$m. Amorphous and crystalline water ice are easily distinguishable at temperatures $\lesssim170~K$: crystalline ice exhibits a narrow feature at 1.65 $\mu$m which is absent in amorphous ice. The feature is suppressed as temperature increases \citep{Grundy1998}. Studying the relative strength and shape of the water ice features in the coma, can yield important information on the impurity, physical structure and the temperature of ice grains. In turn, we may be able to set constraints on the processes experienced by 2I and its formation conditions.

The objectives of our study are: 1) to search for diagnostic absorption features in 2I's coma or on its surface of water ice and other compounds (such as hydrated minerals or organic materials); 2) to compare water ice features in 2I to other Solar system icy objects (e.g., comets, Centaurs and Kuiper belt objects) and set constrains on their birth conditions; 3) to search for the ultra-red organic materials that may be present on the surface or in the coma of 2I as a result of long-term exposure to galactic cosmic rays or UV bombardment. 

\section{Observations and Data Reduction} 
\label{sec:observations}
We observed 2I on three separate occasions. The first two observations were made using director discretionary time allocations. The first observation was carried out in the morning twilight on 2019 September 19 UT using the SpeX spectrograph at the 3-m IRTF telescope. The second observation was made on September 24 UT with the GNIRS spectrograph on the 8-m GEMINI telescope. The last observation was made in service mode again with the SpeX/IRTF on October 9. The observing geometry is shown in Table~\ref{table:obs}.  

The SpeX observations were made with the high throughput prism mode (0.8 -- 2.5 $\mu$m) and a 0.8$^{\prime\prime} \times $15$^{\prime\prime}$ slit that provides a spectral resolving power of $R \sim$ 150 \citep{rayner:2003}. The GNIRS observations were taken with the short camera (plate scale: 0.15$^{\prime\prime}$/pix). The SXD cross-dispersed mode, 32 l/mm grating and 1.0$^{\prime\prime}$ slit, provides an averaged resolving power of R $\sim$ 1700 over 0.9-2.5 $\mu$m \citep{Elias2006}. At least one nearby G-type star was observed together with the comet during each run, which was used both as the telluric correction standard stars and solar analogs. The SpeX data were reduced using the reduction pipeline \texttt{SpeXtool} \citep{cushing:2004}. The GNIRS data were reduced using the Image Reduction and Analysis Facility (IRAF) software and the Gemini IRAF package. Taking the advantage of the similarity between the SpeX data and GNIRS data, we used SpeXtool's \texttt{xtellcor} routine for telluric correction and the \texttt{xmergeorders} routine for merging the GNIRS SXD spectra.

\begin{center}
\begin{table}[h]
\caption{Journal of the 2I Observations. }            
\label{table:obs}      
\begin{tabular}{clccclc}        
\hline\hline                 
UT Date & Facility & $r_h$$^{\dag}$ & $\Delta$$^{\dag}$ & $\alpha$$^{\ddag}$ & $\chi$$^{\S}$ & Standard \\ 
2019    &     &  [au] &     [au] &    [deg] &        &     \\     
\hline                        
Sep 19 & IRTF   & 2.67 & 3.26 & 15.78 & 1.95 & HD283886  \\
Sep 24 & Gemini & 2.60 & 3.16 & 16.77 & 2.15 & SAO79667 \\
Oct 09 & IRTF   & 2.41 & 2.84 & 19.78 & 1.77 & HD87680  \\
\hline                          
\end{tabular}
$^{\dag}$ $r_h$ and $\Delta$ are the heliocentric and geocentric distances, respectively; 
$^{\ddag}$ $\alpha$ is the phase angle;
$^{\S}$ $\chi$ is the airmass.
\end{table}
\end{center}


\section{Results}
\label{sec:analysis}
\begin{figure}[h]
\centering
\includegraphics[angle=00,width=\hsize]{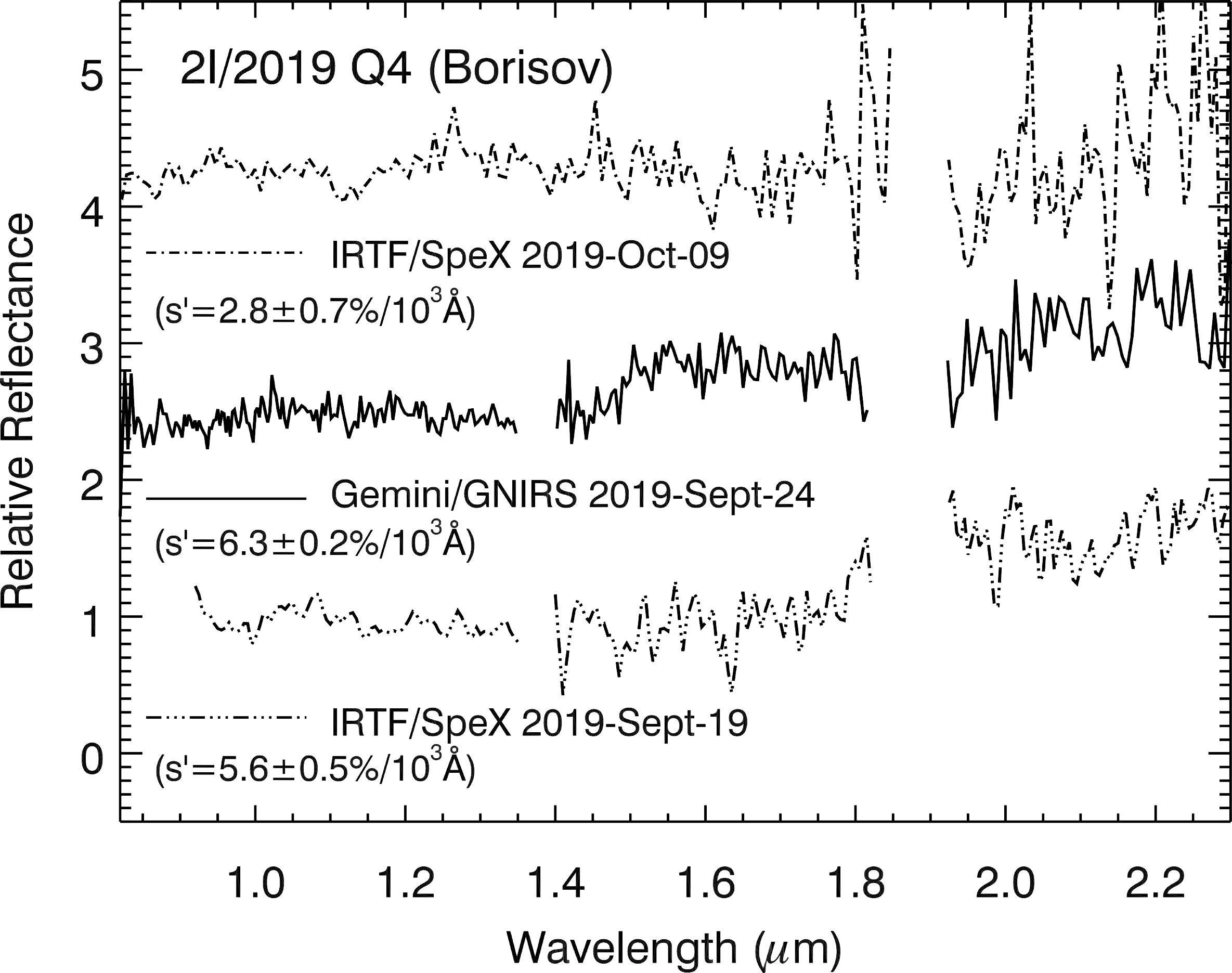}
\caption{NIR spectra of 2I/Borisov, taken with the Gemini-N and IRTF telescopes. The wavelength regions that are severely affected by the atmospheric absorption are omitted. }
\label{fig1}
\end{figure}

The NIR observations of 2I are presented in Figure \ref{fig1}. Although the three spectra were taken on different dates and with different instruments, they consistently appear featureless and show moderately red spectral slopes. We did not detect the diagnostic absorption bands of water ice at 1.5 and 2.0 $\mu$m, respectively. Among the three NIR observations, the Gemini dataset has the highest signal-to-noise ratio (S/N). Thus, we use the Gemini spectra for further analysis.

\subsection{Upper Limits on the Water Ice Abundance}
Previous water ice detections in the cometary comae show that icy grains consist of sub-micron to micron-sized fine particles \citep{Protopapa:2014, yang2014}. However, optical observations suggest that the coma of 2I is dominated by large particles. For example, \cite{jewitt2019} and \cite{Ye2019} consistently suggest that the effective size of the dust particles is about 100$\mu$m. To estimate an upper limit on the amount of possible ice grains in the coma of 2I, we synthesized a pure water ice spectrum using an effective particle radius of 1$\mu$m, 10$\mu$m and 100$\mu$m, respectively, based on the Hapke model \citep{Hapke2012}. We adopted the optical constants of crystalline ice at 150K from \citet{Mastrapa:2008}. We then created an ice mixture using amorphous carbon as a proxy for coma dust. The carbon spectrum is generate the same way as the ice spectrum, using an effective particle radius of 100$\mu$m and the optical constants from \citet{Rouleau1991}. Focusing on the water ice absorption features, we used a low order polynomial function to fit and remove the continuum of the ice mixture as well as the continuum of the spectrum of 2I. Since 2I was observed at high airmass $>$2.0, the spectral regions that are heavily affected by the telluric absorption centered at 1.4 and 1.9 $\mu$m are excluded when fitting the ice features. Studies of Solar system comets have shown that the 1.5$\mu$m band in the comet spectra is vulnerable to impurity and particle size effects and can be absent \citep{yang2009, yang2014}. Therefore, we used the 2-$\mu$m region to estimate water ice abundance, although the S/N is much lower at longer wavelengths, especially in the K-band. Using a least-squares-fitting method around 2-$\mu$m, we estimate that water ice abundance in the coma of 2I is no more than 10\%. 

\begin{figure}[!h]
\centering
\includegraphics[angle=00,width=\hsize]{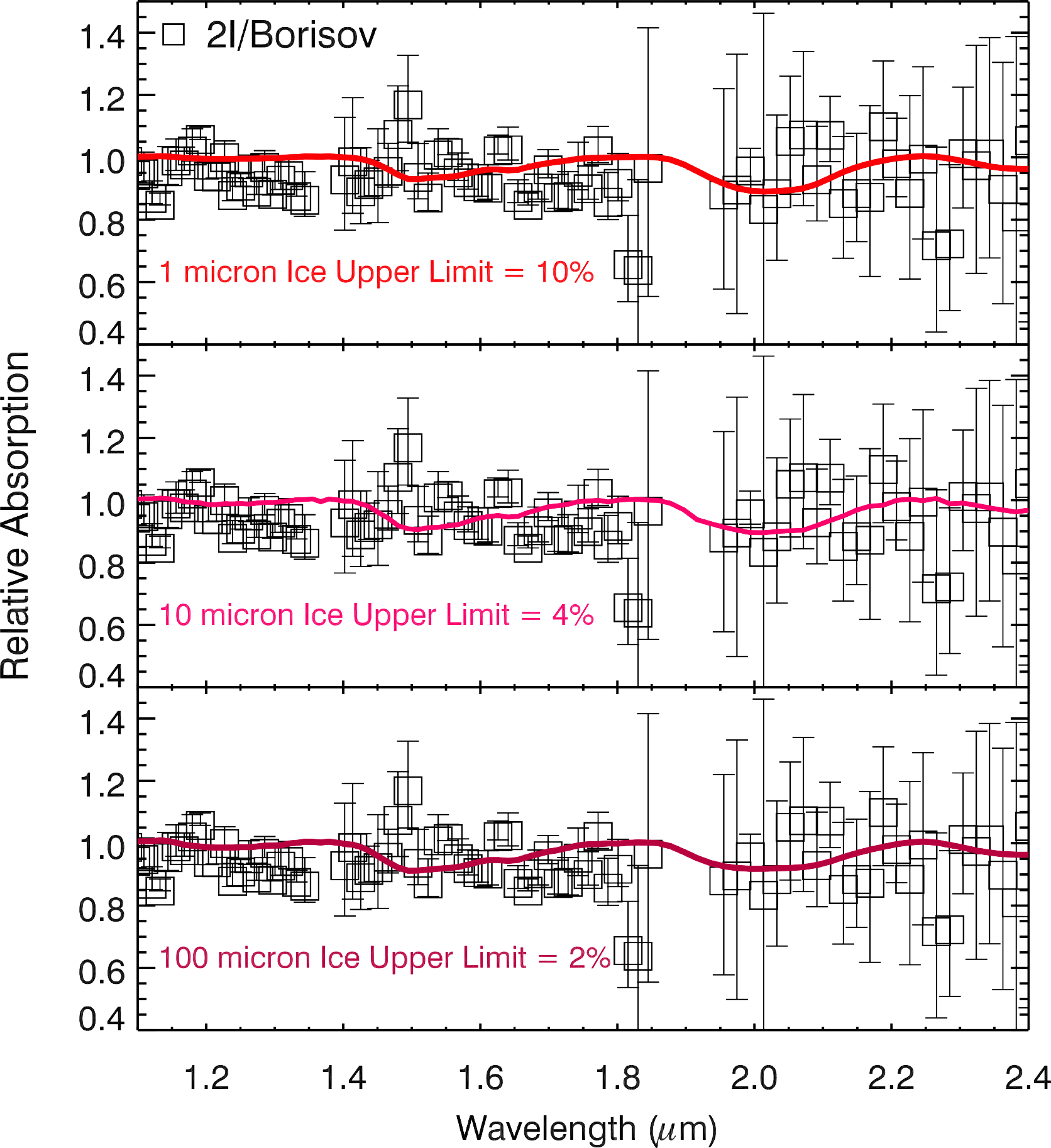}
\caption{The Gemini spectrum of 2I/Borisov obtained on 2019 September 24 UT has been re-sampled to a resolution of $\sim$ 60 (open squares), binning by a factor of 30 and the continuum of the 2I spectrum has been removed, using a low order polynomial model. The red lines are the modeled water ice spectrum using an average particle size of 1$\mu$m, 10 $\mu$m and 100 $\mu$m, respectively. Our model suggests that water ice is present in the coma is no more than 10\%.}
\label{fig2}
\end{figure}

\subsection{Comparison to other objects}
\label{sec:Comparison}
 
\begin{figure}[!h]
\centering
\includegraphics[angle=00,width=\hsize]{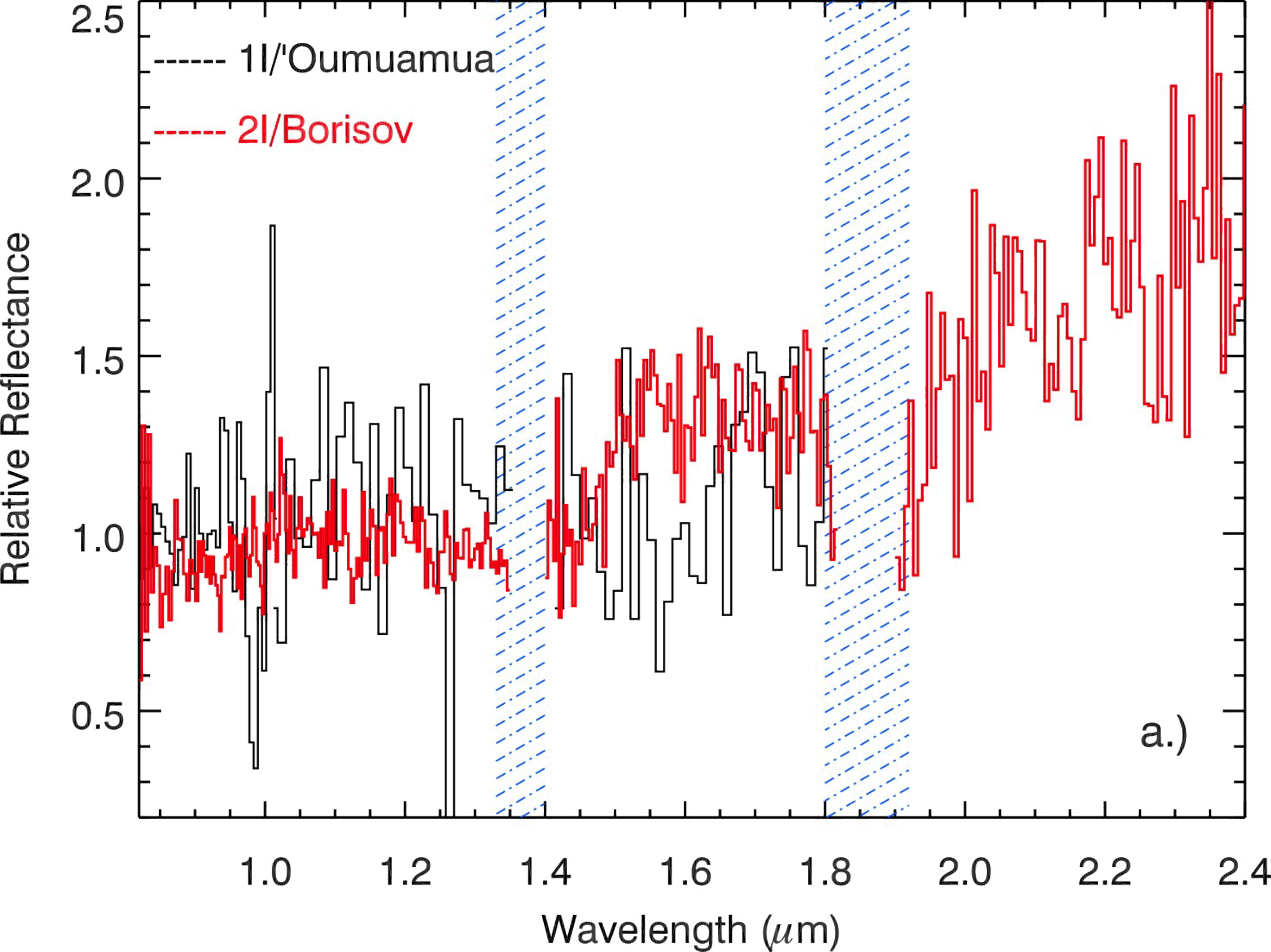}
\includegraphics[angle=90,width=\hsize]{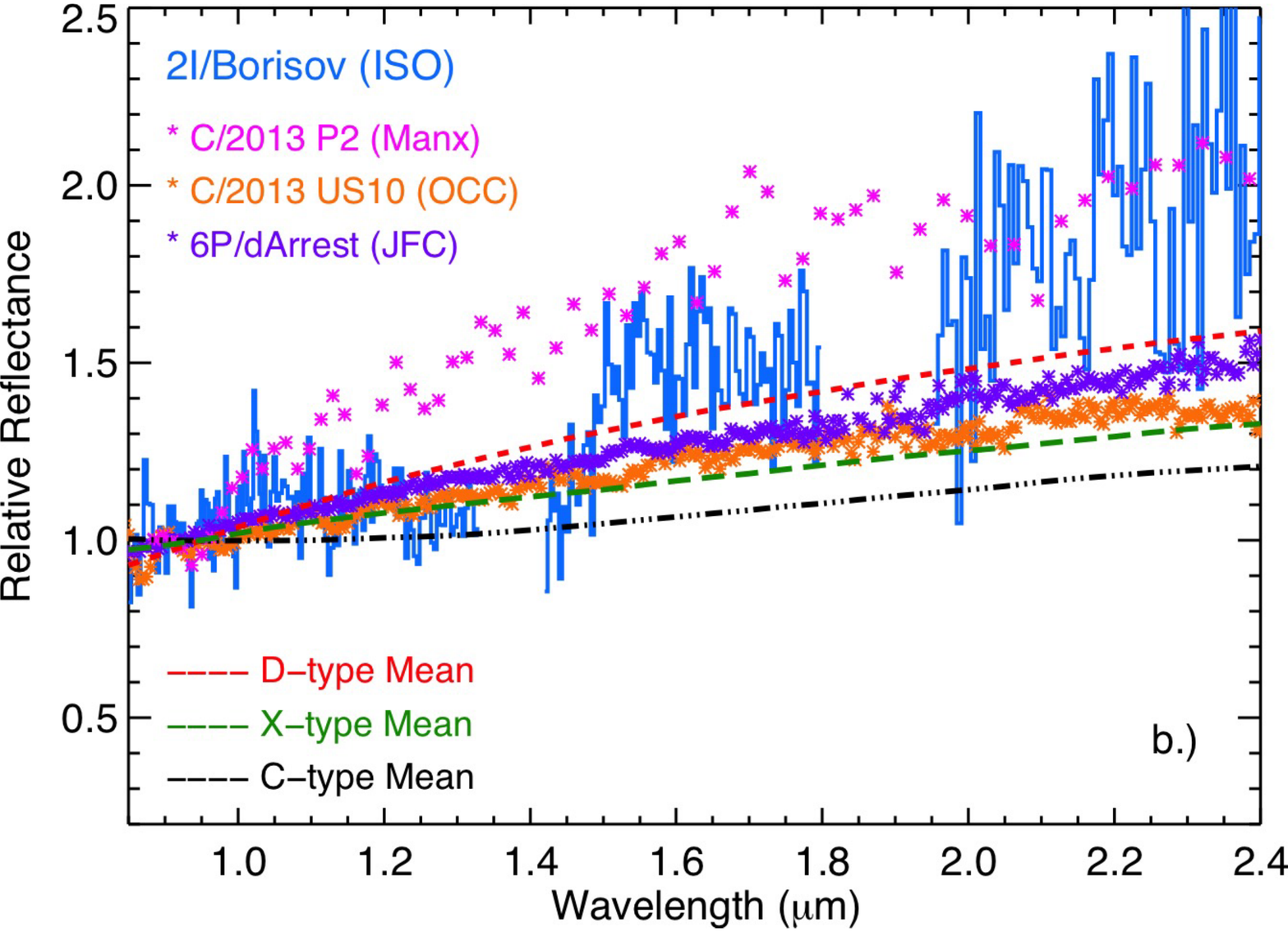}
\caption{(a) Upper: NIR spectrum of 2I/2019 Q4 (Borisov) in comparison with NIR spectrum of 1I/'Oumuamua, taken from \citet{Fitzsimmons2018}. Both ISOs appear featureless and the spectral slopes of the two NIR spectra are comparable. (b) Lower: Comparison between the spectrum of 2I and those of Solar System comets and asteroids. The spectrum of C/2013 US10 is taken from \cite{Protopapa2018}, the spectrum of 6P/d'Arrest and is from \cite{yang2009} and three asteroid spectral classes are taken from \cite{DeMeo2009}. }
\label{fig3}
\end{figure}

The spectra of the two known ISOs are shown in Figure \ref{fig3}a. The spectrum of 2I is similar to that of 1I. However, we note that there was no coma around 1I while the nucleus of 2I is buried inside the coma and not directly detected. Therefore, the spectrum of 2I may not directly reflect the intrinsic composition of the nucleus. At $\lambda < 1.4  \mu m$, 2I shows a red spectral slope that is similar to the mean spectrum of the D-type asteroids. The spectral slope of the 2I spectrum becomes increasingly redder beyond 1.4 $\mu$m and is redder than those of Jupiter family comets (JFC) and the Oort Cloud comets (OCC). Such spectral slope increase may be due to the presence of macromolecular organic materials \citep{Gradie1980}, which become more prominent at longer wavelengths. However, it does not contain the ultra-red material seen in other Solar System bodies \citep{jewitt2019} and is not as red as the recently detected Manx comet C/2013 P2 \citep{Meech2014}, at shorter wavelengths. 

\section{Discussion}
\label{sec:discuss}
Water ice is a key component of the nuclei of Solar system comets and the primary driver of their activity. However, to date, water ice grains in comets (using ground-based telescopes) have only been successfully detected for a limited number of comets, including 17P/Holmes which showed strong water ice features at 2.0 and 3.0 $\mu$m after a major outburst\citep{yang2009}. Based on a survey of 29 individual comets between 1 au and 6 au from the Sun, \cite{Protopapa2018} found that it is challenging to detect solid water ice features in quiescent comets when the comet is close to or within 2.5 au from the Sun due to limited grain lifetime. For instance, strong water ice bands were detected when C/2013 US10 (OCC) was beyond 3.9 au from the Sun and the water ice features disappeared completely even in the 3$\mu$m region when the comet was 2.3 au from the Sun \citep{Protopapa2018}. The only exception to detecting ice features at a small heliocentric distance is the detection of water ice grains in the innermost coma of comet 103P/Hartely 2 (JFC) by the Deep Impact eXtended Investigation \citep{Protopapa:2014}. The well-defined 1.5-, 2.0- and 3.0-$\mu$m water ice absorption features were detected in the spectrum taken 10 km from the nucleus of 103P, when the comet was at 1.06 au from the Sun \citep{Protopapa2014}. There were two populations of water ice grains detected in the coma of 103P, nearly pure water ice with sizes $<$ 5 $\mu$m, and a more distant component of cm-sized icy grains \citep{kelley2013}. With low ejection speeds, the 1 hour lifetime for pure small ($<$5$\mu$m) ice grains ensured that they were present in sufficient numbers to be detected very close to the nucleus \citep{Protopapa:2014}. Compared to 103P, 2I/Borisov is much further away from the Sun and has a much larger effective observing aperture as seen from Earth ($\Delta$=3.16 au). It is possible that solid ice grains are present in the inner coma, but cannot be detected by the ground-based facilities in a large aperture where the spectrum is dominated by dust.     

\begin{figure}[!h]
\centering
\includegraphics[angle=00,width=\hsize]{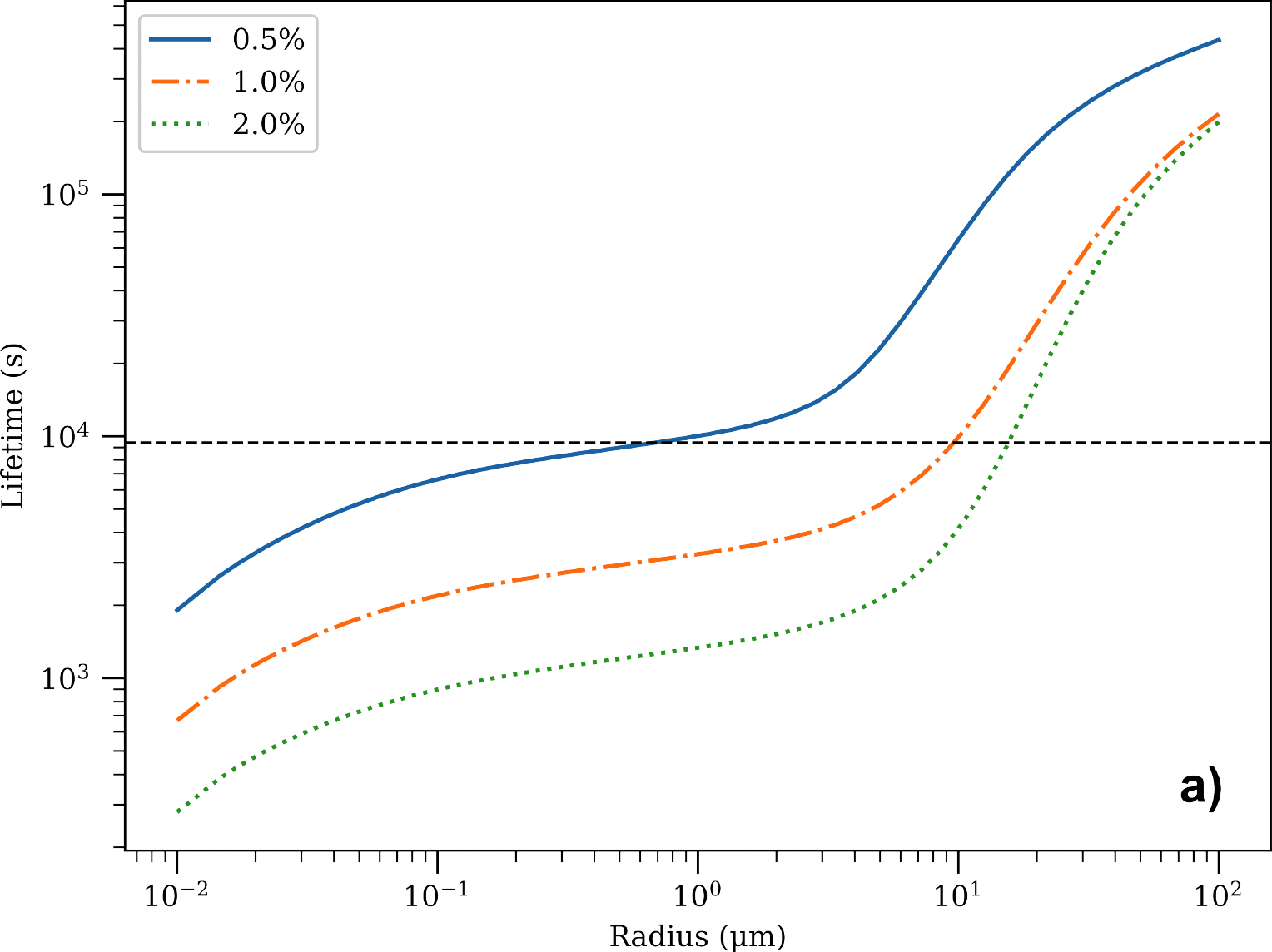}
\includegraphics[angle=00,width=\hsize]{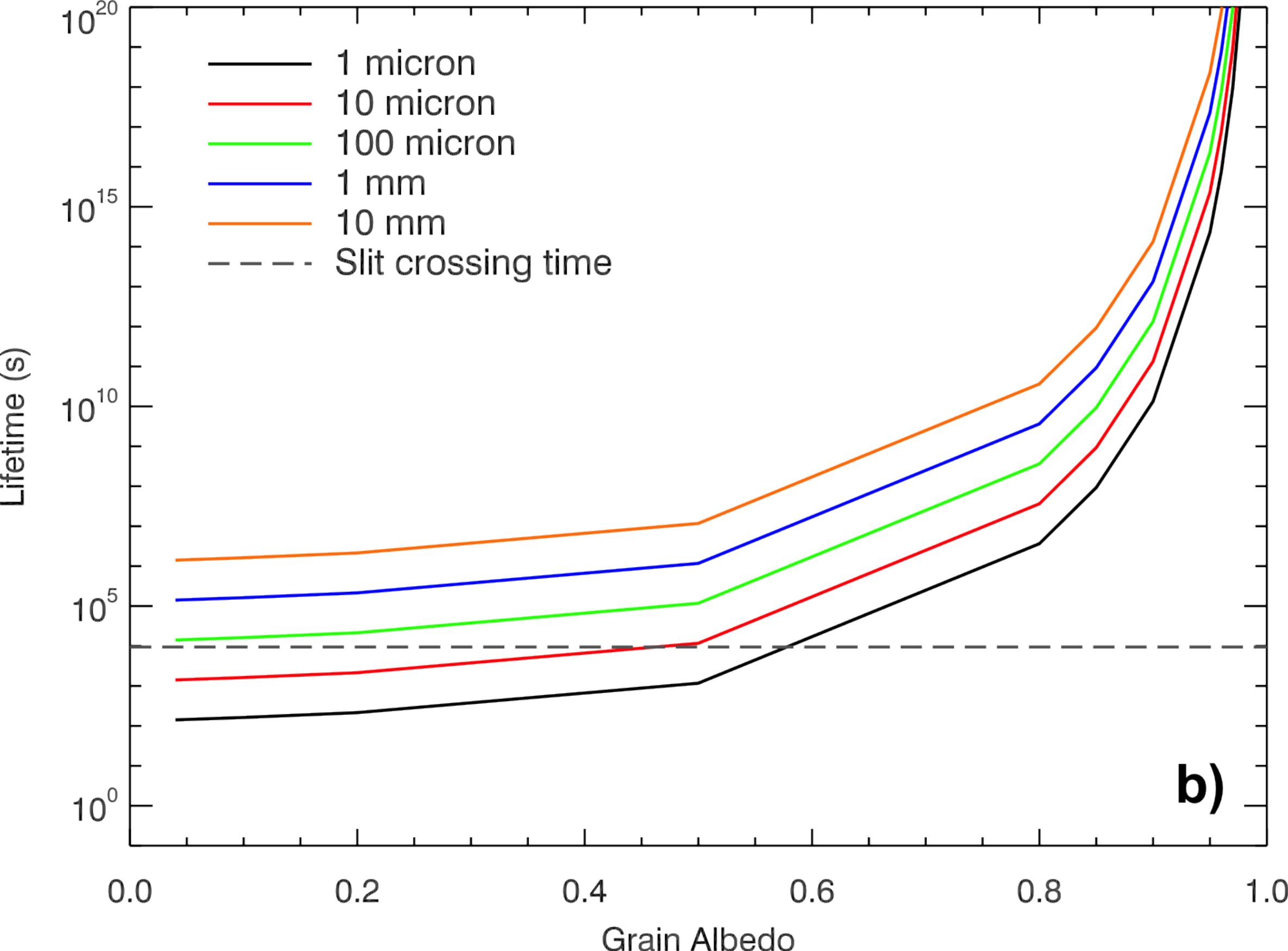}
\caption{Lifetimes of water ice grains as a function of grain radius and geometric albedo at the heliocentric distance of 2.6 au as determined from two independent ice sublimation models. The dashed line indicates the slit crossing time of 9,400s.}
\label{fig4}
\end{figure}

Following the method described in \cite{Protopapa2018}, we estimated the lifetime of water ice grains of various sizes at heliocentric distance around 2.6 au. The balance of energy between absorbed solar radiation, thermal emission, and cooling from sublimation is integrated from a radius of 10~$\mu$m down to 0.01~$\mu$m, at which point a linear decrease is assumed. For small ice grains with radii of 1 $\mu$m, we found the lifetime of 10,000, 3,300 or 1,300s, corresponding to a refractory carbon impurity of 0.5\%, 1.0\% or 2.0\% respectively. For large grains with radii of 10 $\mu$m, the estimated lifetime is 65,000, 9,900, 4,100s for the same fractional impurities. The lifetime for 100 $\mu$m-sized grains with 1.0\% carbon is about  2e5s and a higher fraction of impurity (2\%, 4\%, or 10\%) does not affect the lifetime of such large grains significantly. The extraction aperture for the Gemini spectrum is 1.5$^{\prime\prime}$ (or $\sim$ 2,830 km projected in the plane of the sky). Assuming the expansion velocity of the icy grains is represented by the empirical relation: $v_{dust}$=0.535$\cdot r_h^{-0.6}$ km/s \citep{Whipple1978}, the slit-crossing time is about 9,400s or 2.6 hours. Independently, we modeled the lifetime of icy grains with various albedos and sizes using the method from \citep{meech1986, meech2004}. The derived lifetime of icy grains in comparison to the slit-crossing time is shown in Fig. \ref{fig4}. Since there is no water ice absorption feature observed, we conclude: (1) that 10 $\mu$m radius ice, if present, must have $\gtrsim$ 1\% carbon fraction, (2) that 1 $\mu$m radius ice, if present, must have $\gtrsim$ 0.5\% dirt fraction, or (3) that larger or more pure ice grains are possible, but only if they consist of a small fraction of the coma cross-section and are very close to the nucleus.

As discussed previously in section~\ref{sec:Comparison}, unlike the spectrum of 1I, the spectrum of 2I is not from a bare nucleus but is dominated by the coma surrounding the nucleus. For the case of comet 67P, \cite{Snodgrass2016} noted that the spectral slopes of the coma spectra, obtained with ground-based facilities, are consistent with the values found for the nucleus by Rosetta's Visible, InfraRed and Thermal Imaging Spectrometer. As such, it is not surprising that the two ISOs show comparable spectra and are both lacking the ultra-red matter. \cite{Grundy2009} proposed that mixtures of volatile ice and non-volatile organics could be extremely red and such ultra-red mixtures become progressively darker and less red as the ice sublimates away. It is possible that 2I originally contained red organic materials, which have sublimated or been destroyed by the heat/radiation of the Sun soon after the ISO enters the inner Solar system. Unfortunately, no ice was detected and we cannot constrain the original condition of 2I.  

Shown in Fig. \ref{fig1}, the spectral slope obtained from the third observation is much flatter than that of the earlier observations. Similarly, the decrease in spectral slope for 1I was detected and discussed in \citet{Fitzsimmons2018}, which is attributed to the removal of a surface mantle produced by long-term cosmic ray irradiation. The spectral slope changes as the comet is approaching the Sun may indicate the evolution of organic-rich mixtures in the coma or on the surface of 2I. On the other hand, it is well known that the spectral slope of an active Solar system comet can be affected by the particle sizes in the coma. So the observed slope change in the spectrum of 2I could reflect the change in the particle size distribution in the coma. \cite{Bolin2019} obtained JHK colors of 2I on Sept. 27 and found that 2I exhibits neutral-gray colors in the NIR, which is consistent with our third observation. Since different solar analogs were observed during each observing run, the observed discrepancy in the spectral slope could also be due to the NIR colors of the solar analogs that deviate from that of the Sun. 

\section{Conclusions}
\label{sec:conclusions}
Our near-IR spectrum of 2I/Borisov, taken at the heliocentric distance of $\sim$ 2.6 au, showed no detection of water-ice. 
Based on our observations and analysis, we conclude that:

\begin{enumerate}
\item Large or pure ice grains, if present, comprise no more than 10\% of the coma cross-section and most likely reside in the innermost coma of 2I. 
\item 2I/Borisov displays a moderately red spectrum with a spectral slope of 6\%/1000 \AA\ in the NIR, which is similar to D-type asteroids. 
\item No ultra-red matter was observed, which may either imply that 2I did not contain such material originally or the red material was lost due to the interaction with the Sun.
\end{enumerate}

{\it Acknowledgements}  KJM and JVK acknowledge support through awards from NASA 80NSSC18K0853. BY, MSK and SP are visiting astronomers at the Infrared Telescope Facility, which is operated by the University of Hawaii under contract NNH14CK55B with the National Aeronautics and Space Administration. Based in part on observations obtained at the Gemini Observatory [GN-2019B-Q-DD-102], which is operated by the Association of Universities for Research in Astronomy, Inc., under a cooperative agreement with the NSF on behalf of the Gemini partnership: the National Science Foundation (United States), National Research Council (Canada), CONICYT (Chile), Ministerio de Ciencia, Tecnolog\'{i}a e Innovaci\'{o}n Productiva (Argentina), Minist\'{e}rio da Ci\^{e}ncia, Tecnologia e Inova\c{c}\~{a}o (Brazil), and Korea Astronomy and Space Science Institute (Republic of Korea).
 

\end{document}